\documentclass[aps,prl,reprint,superscriptaddress]{revtex4-1}

\usepackage{amsmath,amssymb}
\usepackage{graphicx}
\graphicspath{{figures/}}
\usepackage[separate-uncertainty=true,multi-part-units=single]{siunitx}
\DeclareSIUnit{\molar}{M}
\renewcommand{\vec}[1]{\mathbf{#1}}
\renewcommand{\tensor}[1]{\mathsf{{#1}}}
\newcommand{\order}[1]{\mathcal{O}\left\{ #1 \right\}}
\newcommand{\abs}[1]{\left\vert #1 \right\vert}

\begin{document}

\title{Trochoidal trajectories of self-propelled Janus particles in a
diverging laser beam} 

\author{Henrique Moyses}
\affiliation{Department of Physics and Center for Soft Matter
  Research, New York University, New York, NY 10003}
 
\author{J\'er\'emie Palacci}
\affiliation{Department of Physics, University of California, San
  Diego, La Jolla, CA 92093}

\author{Stefano Sacanna}
\affiliation{Department of Chemistry and Molecular
    Design Institute, New York University, New York, NY 10003}

\author{David G. Grier}
\affiliation{Department of Physics and Center for Soft Matter
  Research, New York University, New York, NY 10003}

\begin{abstract}
We describe colloidal Janus particles with metallic and
dielectric faces that swim vigorously when illuminated by
defocused optical tweezers without consuming any chemical
fuel.
Rather than wandering randomly, these optically-activated colloidal swimmers
circulate back and forth through the beam of light,
tracing out sinuous rosette patterns.
We propose a model for this mode of light-activated
transport that accounts for the observed behavior through a
combination of
self-thermophoresis and optically-induced torque.
In the deterministic limit, this model yields trajectories that resemble 
rosette curves known as hypotrochoids.
\end{abstract}

\maketitle

\section{Introduction}

Colloidal particles that consume fuel to
translate and rotate are important
examples of active matter \cite{golestanian05,marchetti13}.
Typically taking the form of bifunctional
Janus particles \cite{walther08}, such colloidal
swimmers are driven by gradients of concentration and temperature
that result from chemical reactions \cite{paxton04,golestanian05}.
Several examples of light-activated colloidal swimmers
have been reported whose motions are driven by
thermophoresis
\cite{jiang10,volpe11,buttinoni12,buttinoni13,qian13}.
This mode of motion recently has been analyzed theoretically \cite{bickel14}.
Here, we describe experimental studies of 
the motion of light-activated
colloidal swimmers in nonuniform light fields. 
The interplay of thermophoresis and optical forces in this system
gives rise to highly structured single-particle trajectories
resembling the class of rosette curves known as trochoids.

\subsection{Light-activated swimmers}

\begin{figure}[!t]
 \centering
 \includegraphics[width=0.75\columnwidth]{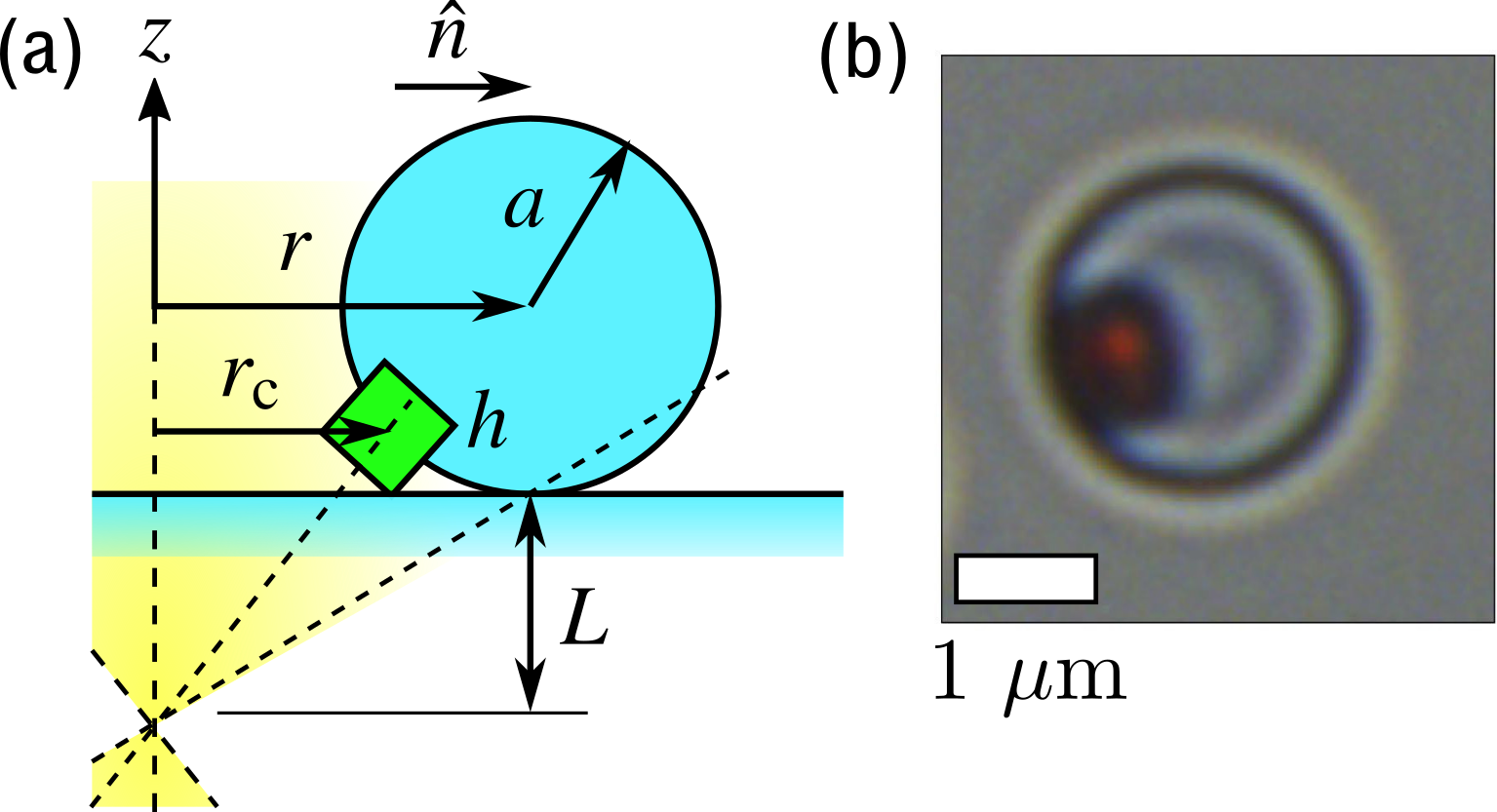}  
 \caption{(a) Schematic representation of a hematite-TPM 
   composite particle being illuminated
   by a diverging laser beam as it rests on a horizontal glass surface.
   (b) Bright-field image of
   a \SI{2}{\um}-diameter swimmer.  Scale bar indicates \SI{1}{\um}.}
  \label{fig:system}
\end{figure}

Our experimental system, shown schematically in 
Figure~\ref{fig:system}(a), consists of a colloidal particle illuminated
by a diverging laser beam.
Each bifunctional particle is comprised of
a \SI{300}{\nm}-wide hematite cube partially embedded in
the surface of a \SI{2}{\um}-diameter
sphere made of 3-methacryloxypropyl trimethoxysilane (TPM)
\cite{sacanna12,abrikosov13,palacci13,palacci14}.
The bright-field microscope image in Figure~\ref{fig:system}(b)
shows a typical particle with its cube visible on the left side.
These particles are dispersed in deionized water at a volume
fraction of \num{e-5}, with \SI{5}{\milli\molar}
tetramethylammonium hydroxide (TMAH) added
to increase the pH to \num{8.5}.
Raising the pH prevents the spheres from sticking to the glass walls
of their container, which is made from a \SI{2}{\cm}-long glass
capillary with cross-section area $\SI{1}{\mm} \times \SI{100}{\um}$
(Vitrocom, catalog number 5010).
Both TPM and hematite are substantially denser than water,
and the composite particles rapidly
sediment to the bottom wall, as depicted
in Figure~\ref{fig:system}(a).

TPM is a transparent dielectric at optical wavelengths.
Hematite, by contrast, absorbs visible light strongly.
When a composite particle is illuminated, its hematite side becomes warmer
than its TPM side.
The local temperature gradient gives rise to thermophoresis,
which causes the particle to move
\cite{weinert08,jiang10,yang13}.
In uniform illumination, this motion is essentially ballistic,
with rotational diffusion causing small deviations in a
swimmer's trajectory \cite{jiang10,bickel14}.
Here, we show that nonuniform illumination engenders
motion of a very different nature, with the particle tracing
out continuous loops that pass through the intensity maximum
and turn around near the periphery of the light field.
This behavior differs from previous reports of thermophoretic
swimmers in optical traps \cite{jiang10} whose trajectories
appeared diffusive.

\begin{figure*}[!t]
 \centering
 \includegraphics[width=0.9\textwidth]{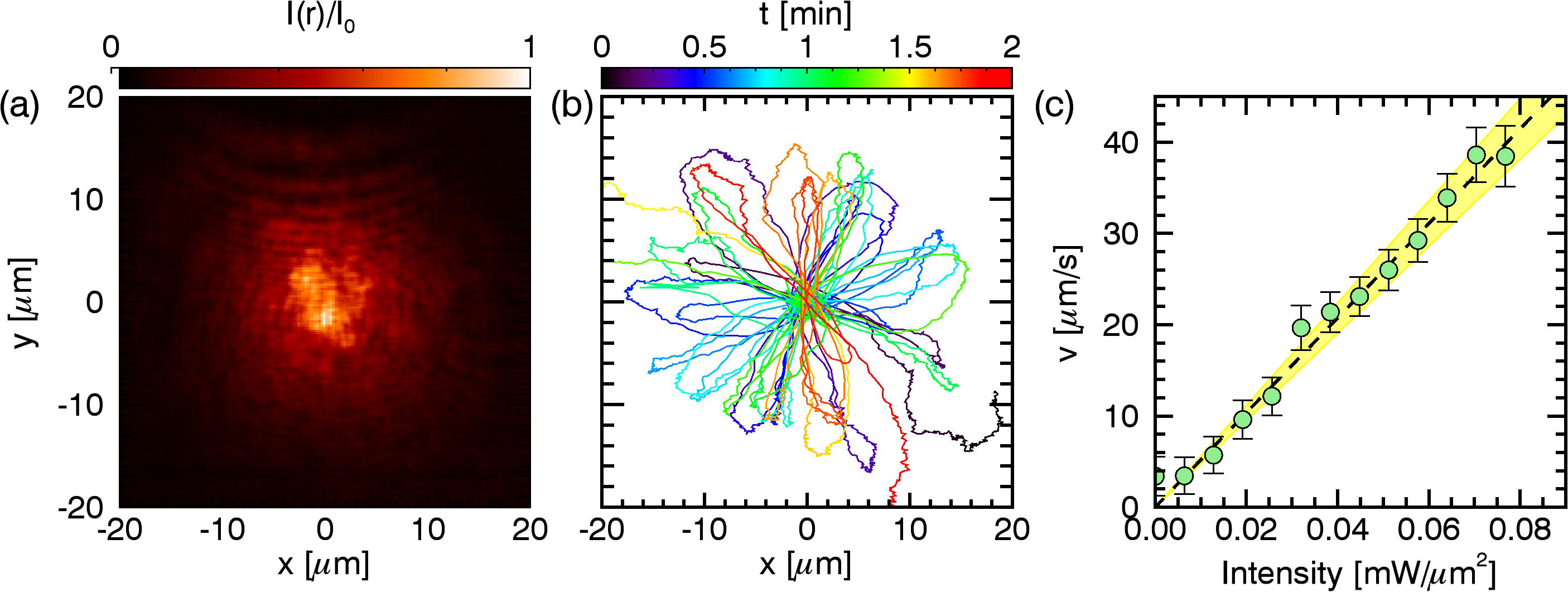}  
 \caption{(a) Measured intensity distribution $I(\vec{r})$
   of laser illumination in the plane of the swimmer's motion.
   The color bar indicates values relative to the peak value,
   $I_0$, at the center.
   (b) Two minutes of a typical 
   swimmer's trajectory colored by time.  The trajectory loops
   repeatedly through the center of the intensity distribution.
   (c) Relationship between the swimming speed, $v(\vec{r})$,
   and the local light intensity, $I(\vec{r})$. The dashed line is a
   one-parameter fit for the associated mobility,
   $\alpha = \SI{518(41)}{\cubic\um\per\second\per\milli\watt}$.}
  \label{fig:light_trajectory}
\end{figure*}

\subsection{Illumination and imaging}

Our instrument consists of an inverted holographic
optical trapping system
that operates at a vacuum wavelength of \SI{532}{\nm}
(Coherent Verdi) and creates patterns of optical tweezers with
computer-generated holograms that are imprinted onto the laser
beam with a liquid-crystal spatial light modulator 
(Hamamatsu X13267-04 LCoS SLM).
These holograms are projected into the sample 
using a microscope objective lens (Nikon Plan-Apo $60\times$,
numerical aperture 1.4, oil immersion).
The single trap used for this study was powered with
\SI{0.6}{\watt} at the laser's output, which corresponds
to roughly \SI{0.1}{\watt} delivered to the sample.

The same objective lens is used to relay images of swimmers
interacting with the optical traps onto an integrated
video camera (NEC TI-324A), which records them
at even time intervals of $\tau = \SI{33.3}{\milli\second}$.
The microscope's
$\SI{90}{\um} \times \SI{70}{\um}$ field of view typically
contains one particle.
Images are analyzed with standard 
methods of digital video microscopy \cite{crocker96,krishnatreya14a}
to measure a swimmer's trajectory, $\vec{r}_n = \vec{r}(t_n)$
at times $t_n = n \tau$ that are integer multiples of the frame interval.
Because the particle remains sedimented onto the lower glass
surface with its hematite cube inclined downward, we can 
define its orientation $\hat{n}(t)$ to be directed from
the center of the cube to the center of the sphere
along the horizontal plane.

The illuminated particle
swims in the direction of $\hat{n}(t)$, which is to say away from
its hotter end.
This differs from these particles' motion in the presence
of hydrogen peroxide \cite{palacci13,palacci14}.
In this case, decomposition of the chemical fuel engenders
osmotic flows that propel the particle in the opposite direction.
In both cases, the phoretic flow responsible for propulsion
does not lift the swimmers off the lower
glass surface; their motions are essential two-dimensional. 

Rather than focusing an optical trap onto such a particle,
we instead focus it $L \approx\SI{10}{\um}$ 
below the lower glass surface, as indicated in
Figure~\ref{fig:system}(a),
so that the colloidal particle is illuminated by a diverging 
beam that propagates upward and outward.
Figure~\ref{fig:light_trajectory}(a) shows the intensity
distribution, $I(\vec{r})$, in
the plane of the particle that is measured by replacing
the sample with
a front-surface mirror in the same plane.
This distribution is well described by a Gaussian surface
of revolution,
\begin{equation}
  \label{eq:intensitydistribution}
  I(\vec{r}) = I_0 \, \exp\left(-\frac{r^2}{2 \sigma^2}\right),
\end{equation}
with width $\sigma = \SI{7.3(2)}{\um}$.

\subsection{Trajectory analysis}

As soon as the particle is illuminated, it starts to translate across
the lower glass wall of its container.
The data in Figure~\ref{fig:light_trajectory}(b) show a typical
\SI{2}{\minute} trajectory.
The swimmer's trajectory is localized around the center of brightness
and has a looping character that is quite distinct from Brownian
motion.
The trajectory curves gently as the swimmer passes through the
center of the laser beam, and loops back around at the beam's
periphery to create a rosette pattern.
Thermal randomness is evident in the trajectory's small deviations.
The coherent large-scale motion, however, suggests the action of an
underlying deterministic process.
Comparable behavior is observed for all particles in the sample;
the example in Figure~\ref{fig:light_trajectory}(b) is typical.
This motion continues for several minutes at a time, and ends when
the particle eventually moves out of the light and stops moving altogether.

We estimate the swimmer's instantaneous velocity from the measured trajectory as
\begin{equation}
  \label{eq:measuredvelocity}
  \vec{v}_n = \frac{\vec{r}_n - \vec{r}_{n-1}}{\tau}
\end{equation}
at the mid-point position
\begin{equation}
  \label{eq:measuredposition}
  \bar{\vec{r}}_n = \frac{1}{2} (\vec{r}_n + \vec{r}_{n-1}).
\end{equation}
The trajectory-averaged position dependence of the swimmer's
speed then can be computed from the $N$-point trajectory
using an adaptive kernel density estimator
\cite{silverman92},
\begin{equation}
  \label{eq:speedestimator}
  v(\vec{r})
  =
  \frac{1}{\sqrt{2} (N-1) \rho(\vec{r})}
  \sum_{n = 2}^N \frac{v_n}{\sigma_n} \, 
  \exp\left(
    -\frac{\abs{\vec{r} - \bar{\vec{r}}_n}^2}{2\sigma_n^2}
  \right),
\end{equation}
whose width $\sigma_n$ is
selected based on the density of measurements.
The speed estimate is normalized by the density of measurements,
\begin{equation}
  \label{eq:densityestimator}
  \rho(\vec{r})
  =
  \frac{1}{\sqrt{2}(N-1)} \sum_{n = 2}^N 
  \frac{1}{\sigma_n}
  \exp\left(
    \frac{\abs{\vec{r} - \bar{\vec{r}}_n}^2}{2\sigma_n^2}
  \right).
\end{equation}
The data in Figure~\ref{fig:light_trajectory}(c)
show how the swimming speed depends on the local
light intensity obtained by combining results for $v(\vec{r})$ with 
those for $I(\vec{r})$.
The swimmer moves most rapidly as it passes through the
center of the beam, where the light is brightest.

\section{Self-thermophoretic swimming} 
\label{sec:thermophoresis}

We propose that the swimmer moves through the light field primarily
by self-thermophoresis engendered by optically-induced heating.
Specifically, the hematite cube's temperature rises as it 
absorbs light from the beam.
The resulting temperature gradient, $\nabla T(\vec{r})$, 
then induces motion through
thermophoresis with a drift velocity,
\begin{equation}
  \label{eq:thermophoresis}
  \vec{v}(\vec{r}) = D_T \, \nabla T(\vec{r}).
\end{equation}
The thermal diffusion coefficient, $D_T$, depends on the properties
of the species dissolved in the solvent \cite{iacopini06,yang13}, and 
presumably is dominated by TMAH.
When dispersed in solutions
without TMAH, these particles still move under laser illumination, but at
only a tenth of the speed.
The resulting trajectories are far more strongly influenced
by both rotational
and translational Brownian motion and lack the looping
nature of the trajectory plotted in Figure~\ref{fig:light_trajectory}(b).
This is consistent with previous studies of bulk thermophoresis
in which added salt is found to play a similar role \cite{fayolle08}.

Unlike previous studies of particle motion in externally-imposed
temperature gradients \cite{yang13}, the temperature gradient
in this system is generated at the position of the particle
through optically-induced heating of the hematite cube.
Both $T(\vec{r})$ and $\nabla T(\vec{r})$ vary as the particle
moves through $I(\vec{r})$.
Previous reports
suggest that $D_T$ may be largely independent of the temperature
\cite{iacopini06,yang13}.
In that case, we might expect the swimmer's translation speed
$v(\vec{r})$ to be proportional to the local intensity,
$I(\vec{r})$.
A straightforward model for the swimmer's motion is therefore
\begin{equation}
  \label{eq:motionmodel}
  \vec{v}(\vec{r}) = \alpha \, I(\vec{r}) \, \hat{n},
\end{equation}
where $\hat{n}$ denotes the orientation of the swimmer
from the center of the cube to center of the sphere,
as indicated in Figure~\ref{fig:system}(a).
The dashed line superimposed on the data in
Figure~\ref{fig:light_trajectory}(c)
is a one-parameter fit to Eq.~\eqref{eq:motionmodel}
with $\alpha = \SI{518(41)}{\cubic{\um}\per\second\,\milli\watt}$.
Without a direct probe of the local temperature, however, the
relationship between $I(\vec{r})$ and $\nabla T(\vec{r})$
is difficult to gauge, which precludes 
using $\alpha$ to estimate $D_T$.

\subsection{Flow field generated by a stationary swimmer}
\label{sec:swimmerflow}

We test this model for self-thermophoretic swimming by using
tracer particles to map the flow field created by a stationary swimmer
that is affixed to the lower glass surface.  We then compare the
measured velocity field to predictions of a hydrodynamic model 
that accounts for the tracer particles' own thermophoresis.
Correcting for the tracers' thermophoresis yields an estimate for
the stationary swimmer's intrinsic flow field.
This, in turn, is related to the hydrodynamic forces that enable
a free particle to swim.

The external force that prevents the stuck swimmer from moving
is transferred to the fluid, thereby generating a flow field.
Because we are primarily interested in the far-field flow profile,
we model this force as a point source of flow,
or stokeslet, located at the center of the swimmer and directed
opposite to the orientation vector, $\hat{n}$.
The Oseen tensor for this flow is \cite{pozrikidis92}
\begin{equation}
  \label{eq:stokeslet}
  \tensor{G}_{\alpha\beta}^S(\vec{r})
  =
  \frac{1}{8 \pi \eta r} \,
  \left(
    \delta_{\alpha\beta} + \frac{r_\alpha r _\beta}{r^2}
  \right),
\end{equation}
where $\eta = \SI{e-3}{\pascal\second}$ is the viscosity of water.
The flow at position $\vec{r}$ due to a force $\vec{f}(\vec{r}_0)$
acting on a particle at
$\vec{r}_0$ is
\begin{equation}
  \label{eq:stokesletflow}
  \vec{u}^S(\vec{r})
  =
  \tensor{G}^S(\vec{r} - \vec{r}_0) \cdot \vec{f}(\vec{r}_0) .
\end{equation}
For the particular case of light-induced self-thermophoresis,
the driving force acting on the fluid should be comparable
to the force responsible for the free swimmer's motion,
\begin{subequations}
  \label{eq:drivingforce}
\begin{align}
  \vec{f}(\vec{r}_0) 
  & \approx 
  \frac{1}{\mu} \, \vec{v}(\vec{r}_0) \\
  & = 
  \frac{1}{\mu} \, \alpha \, I(\vec{r}_0) \, \hat{n},
\end{align}
\end{subequations}
where $\mu$ is the swimmer's wall-corrected mobility.
We further assume that this point force operates on the fluid
at the center of the sphere.

To satisfy no-slip boundary conditions at the nearby
glass surface, we adopt the method of images
\cite{blake71,pozrikidis92}, in which the flow at the surface
is identically canceled by 
the source's hydrodynamic image in the surface.
The hydrodynamic image for the stokeslet at $\vec{r}_0$
consists of a stokeslet
located at the mirror position, $\vec{r}_1 = \vec{r}_0 - 2 a \hat{z}$,
and oriented in the opposite direction, plus additional
contributions \cite{blake71}
from a source doublet (SD), 
\begin{equation}
  \label{eq:sourcedoublet}
  \tensor{G}_{\alpha\beta}^{SD}(\vec{r}) 
  =
  \frac{1}{8 \pi \eta r^3} 
  \left(
    - \delta_{\alpha\beta}
    + 3 \frac{r_\alpha r_\beta}{r^2}
  \right),
\end{equation}
and a stokeslet dipole (D) oriented in direction 
$\hat{d} = \hat{n} - 2 n_z \hat{z}$,
\begin{equation}
  \label{eq:stokesletdipole}
  \tensor{G}_{\alpha\beta}^D(\vec{r}, \hat{d})
  =
  \frac{1}{8 \pi \eta r^3}
  \biggl[
    \hat{d} \cdot \vec{r}
    \left(
      \delta_{\alpha\beta} 
      + 3 \frac{r_\alpha r_\beta}{r^2}
    \right) 
    - (r_\alpha d_\beta + r_\beta d_\alpha)
  \biggr].
\end{equation}
The flow field generated by this image system is
\begin{subequations}
\begin{align}
  \label{eq:imageflow}
  \vec{u}^I(\vec{r})
  & =
    -\tensor{G}^S(\vec{r}-\vec{r}_1) \cdot \vec{f}(\vec{r}_0)
    - 2 a^2 \, \tensor{G}^{SD}(\vec{r} - \vec{r}_1) \cdot \hat{d} \,
    f(\vec{r}_0)  \nonumber \\
  & \quad \quad
    + 2 a \, \tensor{G}^D(\vec{r} - \vec{r}_1,\hat{d}) \cdot \hat{z}
    \, \vec{f}(\vec{r}_0) \cdot \hat{z} \\
  & =
    \tensor{G}^I(\vec{r} - \vec{r}_1) \cdot \vec{f}(\vec{r}_0).
\end{align}
\end{subequations}
The total Oseen tensor for this model is then
\begin{equation}
  \label{eq:totaloseentensor}
  \tensor{G}(\vec{r} - \vec{r}_0) 
  = 
  \tensor{G}^\text{S}(\vec{r} - \vec{r}_0) 
  + 
  \tensor{G}^\text{I}(\vec{r} - \vec{r}_1),
\end{equation}
and the total flow at position $\vec{r}$ is
\begin{equation}
  \label{eq:selfthermophoreticflowfield}
  \vec{u}(\vec{r}) = -\alpha I(\vec{r}_c) \, \tensor{G}(\vec{r}-\vec{r}_0) \cdot \hat{n}.
\end{equation}

We neglect hydrodynamic coupling to 
the more distant parallel wall,
which would add substantially to the complexity
of the model \cite{liron76}, but would not appreciably
influence particles' motions near the lower wall
\cite{dufresne01}.

Were the particle free to move, the reaction
force, $-\vec{f}(\vec{r}_0)$, would give rise to a stresslet flow
associated with force-free motion \cite{spagnolie12}.
In the absence of the external restraining force,
the swimmer then would move with velocity
\begin{equation}
  \label{eq:swimmervelocity}
  \vec{v}(\vec{r}) = - \vec{u}(\vec{r}).
\end{equation}

\subsection{Accounting for tracer thermophoresis}
\label{sec:tracers}

\begin{figure}[!t]
  \centering
  \includegraphics[width=0.65\columnwidth]{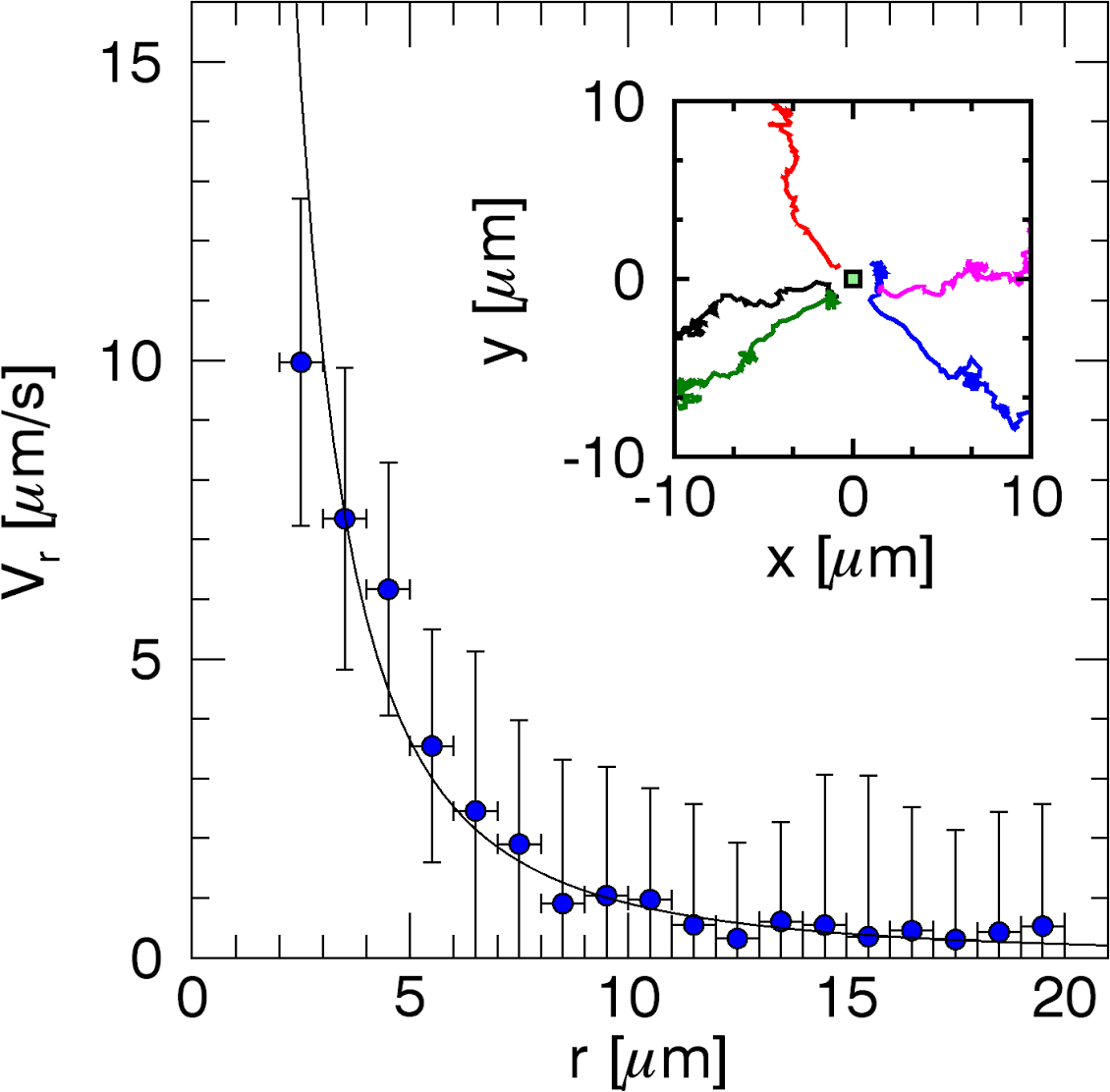}    
  \caption{The measured radial speed, $V_r(r)$, of tracer
    particles approaching an illuminated hematite cube at $r = 0$, 
    computed according to Eq.~\eqref{eq:tracerradialspeed}.
    The solid curve is a one-parameter fit
    to Eq.~\eqref{eq:tracerphoresis}. 
    Inset: Representative trajectories of five tracer spheres drifting
    inward toward the stationary hematite cube.}
  \label{fig:tracerphoresis}
\end{figure}

We map the flow field, $\vec{u}(\vec{r})$, by measuring the
trajectories \cite{crocker96}
of silica spheres of radius $a_{tr} = \SI{150}{\nm}$ that are
dispersed in the system as tracer particles.
Their velocity,
$\vec{V}(\vec{r})$,
in the presence of a stationary swimmer located at $\vec{r} = 0$
provides insight into the swimmer-induced flow field.
According to Fax\'en's first law,
a tracer sphere of radius $a_{tr}$ immersed in
a flow field $\vec{u}(\vec{r})$ and
driven by an external force $\vec{F}(\vec{r})$
has an instantaneous
velocity
\begin{equation}
  \label{eq:faxen}
  \vec{V}(\vec{r}) 
  = 
  \vec{u}(\vec{r}) 
  +
  \frac{a_{tr}^2}{6} \nabla^2 \vec{u}
  +
  \tensor{b}(\vec{r}) \cdot \vec{F}(\vec{r}),
\end{equation}
where $\tensor{b}(\vec{r})$ is the sphere's mobility
tensor.

We include $\vec{F}(\vec{r})$ in Eq.~\eqref{eq:faxen}
to account for thermophoresis of the tracer particles
in the non-uniform temperature field generated by the
hot hematite cube.
The steady-state temperature profile around
the heated cube satisfies the Laplace equation
and therefore may be modeled as 
\begin{equation}
  \label{eq:temperature}
  T(\vec{r}) = T_\infty + \frac{h}{r} (T_c - T_\infty),
\end{equation}
where $T_\infty$ is the temperature far from the cube,
$T_c$ is the cube's temperature, and $h$ is the length
of the cube's side.
Equation~\eqref{eq:temperature} applies
in the far field, for $r > h$.
A tracer particle's thermophoretic velocity therefore should be
\begin{equation}
  \label{eq:tracerphoresis}
  \tensor{b}(\vec{r}) \cdot \vec{F}(\vec{r}) =
  - D_T^{tr} (T_c - T_\infty) h \, \frac{1}{r^2} \hat{r},
\end{equation}
where $D_T^{tr}$ is the tracers' thermal diffusion coefficient.
Thermophoresis causes the tracers to drift inward
toward the cube.

To test this model and to estimate the relevant coefficients,
we measure tracer particles' trajectories in the presence of
bare hematite cubes stuck to the lower glass wall.
This isolates the drift due to tracer thermophoresis
from the flow due to the swimmer's self-thermophoresis.
Figure~\ref{fig:tracerphoresis} shows the ensemble average of
tracers' measured radial speeds,
\begin{equation}
  \label{eq:tracerradialspeed}
  V_r(r) = \frac{1}{2\pi} \int_0^{2\pi} \abs{\vec{V}(\vec{r})
    \cdot \hat{r}} \, d\theta,
\end{equation}
averaged over angles centered on the heated cube.
Typical trajectories are plotted in the inset to Figure~\ref{fig:tracerphoresis}.
The solid curve in Figure~\ref{fig:tracerphoresis} is a one-parameter
fit of $V_r(r)$ to the prediction of Eq.~\eqref{eq:tracerphoresis}, which yields
$D_T^{tr} (T_{c}-T_{\infty})h = \SI{90(10)}{\cubic\um\per\second}$.

This model for tracer thermophoresis omits other possible 
optically-mediated influences
on the tracers' motion
such as advection by thermally-induced convection currents.
Omitting such contributions is justified for the present system by the
observation that Eq.~\eqref{eq:tracerradialspeed}
agrees well with experimental results.

\begin{figure*}[!t]
  \centering
  \includegraphics[width=0.9\textwidth]{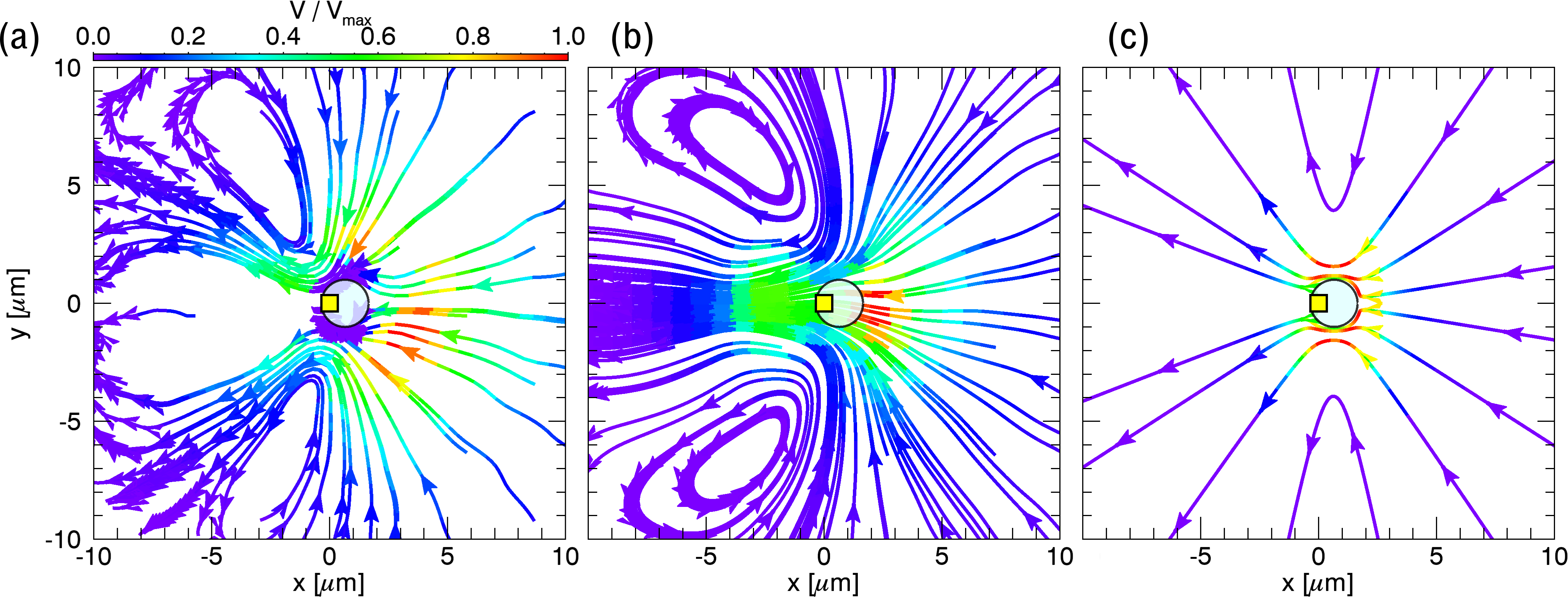}
  \caption{(a) Streamlines of the velocity field, $\vec{V}(\vec{r})$, 
    of silica tracer particles
    moving in the vicinity of a stuck swimmer illuminated 
    by a diverging laser beam.
    (b) Streamlines of the velocity field obtained from 
    simulated trajectories of tracer particles moving under the combined
    influence of a stuck swimmer's flow field and thermophoresis
    in the swimmer's temperature field.
    (c) Streamlines of the swimmer's flow field, $\vec{u}(\vec{r})$,
    used in computing (b).}
  \label{fig:flowfield}
\end{figure*}

\subsection{Other influences on the tracers' motions}
\label{sec:otherinfluences}

The silica tracer spheres also are acted upon directly
by optical forces.
We measure this influence by moving the illumination
into a region without any hematite cubes and tracking nearby spheres' motions.
These measurements reveal that the spheres are
very weakly repelled from the optical axis by the 
diverging beam's radiation pressure.
The maximum radial drift velocity induced by radiation
pressure is much smaller than $\SI{1}{\um\per\second}$
and thus no more than a few percent of the thermophoretic
or flow-induced drift.

Gravity also influences the tracers' motions, causing
them to drift toward the lower wall of the sample chamber.
The resulting sedimentation is slow enough that it may be ignored without
qualitatively influencing the observed velocity field.

\subsection{Mapping a stationary swimmer's flow field}
\label{sec:flowfield}

Figure~\ref{fig:flowfield}(a) shows streamlines of the tracer-particle
velocity field, $\vec{V}(\vec{r})$, 
under steady illumination.
The position of the static swimmer is indicated by a circle 
plotted to scale near the center of the figure,
with the cube plotted to scale as a square to the left of the swimmer's center.
The velocity field was computed from 250 independent
tracer-particle trajectories with a kernel-density
estimate equivalent to Eq.~\eqref{eq:speedestimator}.
Streamlines were computed with fourth-order Runge-Kutta 
integration starting from random initial positions.
Colors represent the local speed, $V(\vec{r})$, and arrows
indicating the direction of motion are evenly separated in
time.
Tracer particles flow predominantly from right to left in
Figure~\ref{fig:flowfield}, with two clear vortexes appearing
slightly downstream of the swimmer.
No streamlines are plotted directly downstream of the swimmer
because no tracer particles ventured into this region.
Any tracers originally in that region were expelled to the left
before recording began. 

Figure \ref{fig:flowfield}(b) shows streamlines of the
flow field predicted by Fax\'en's law, Eq.~\eqref{eq:faxen},
using as inputs the model for the swimmer's self-thermophoretic 
flow field from Eq.~\eqref{eq:selfthermophoreticflowfield} and the
result for the tracers' thermophoretic drift from
Eq.~\eqref{eq:tracerphoresis}.
Qualitative features of the computed velocity field,
$\vec{V}(\vec{r})$,
agree very well with the experimental result.
The model's success at reproducing the experimentally
observed tracer-particle velocity field justifies its
underlying assumptions and approximations.

Incorporating the influence of radiation pressure and sedimentation
has little influence on the computed results.
Omitting the tracers' thermophoresis, however,
eliminates the paired vortexes downstream of the swimmer.
This can be seen in Figure~\ref{fig:flowfield}(c), which shows
streamlines of the swimmer-generated flow field, $\vec{u}(\vec{r})$,
that is calculated with Eq.~\eqref{eq:selfthermophoreticflowfield} 
and was used to compute Figure~\ref{fig:flowfield}(b).
Tracer thermophoresis therefore qualitatively influences
our observations;
the swimmer's intrinsic flow field is a continuous
stream.
The peak flow speed at the swimmer's position
in Figure~\ref{fig:flowfield}(c)
is consistent with the swimmer's speed when it swims
freely, with no adjustable parameters.

\section{Optical forces and torques}
\label{sec:opticalforces}

Although the self-thermophoretic mechanism
explains the swimmers' overall propulsion, it does not explain
their trajectories' looping character.
Were this the only mechanism influencing the particles' motion,
they would move rapidly to the periphery of the light field, where
they would proceed to diffuse.
Only if their rotational and translational
diffusion directed them back into light would they swim back toward
the axis.  This process would be slow, however, and the particles
would be more likely to diffuse away from the optical axis.
Apparently, another mechanism is at work.
The nature of this additional influence
reveals itself through subtle
biases in the thermophoretically-driven motion.
 
\begin{figure}
 \centering
  \includegraphics[width=0.95\columnwidth]{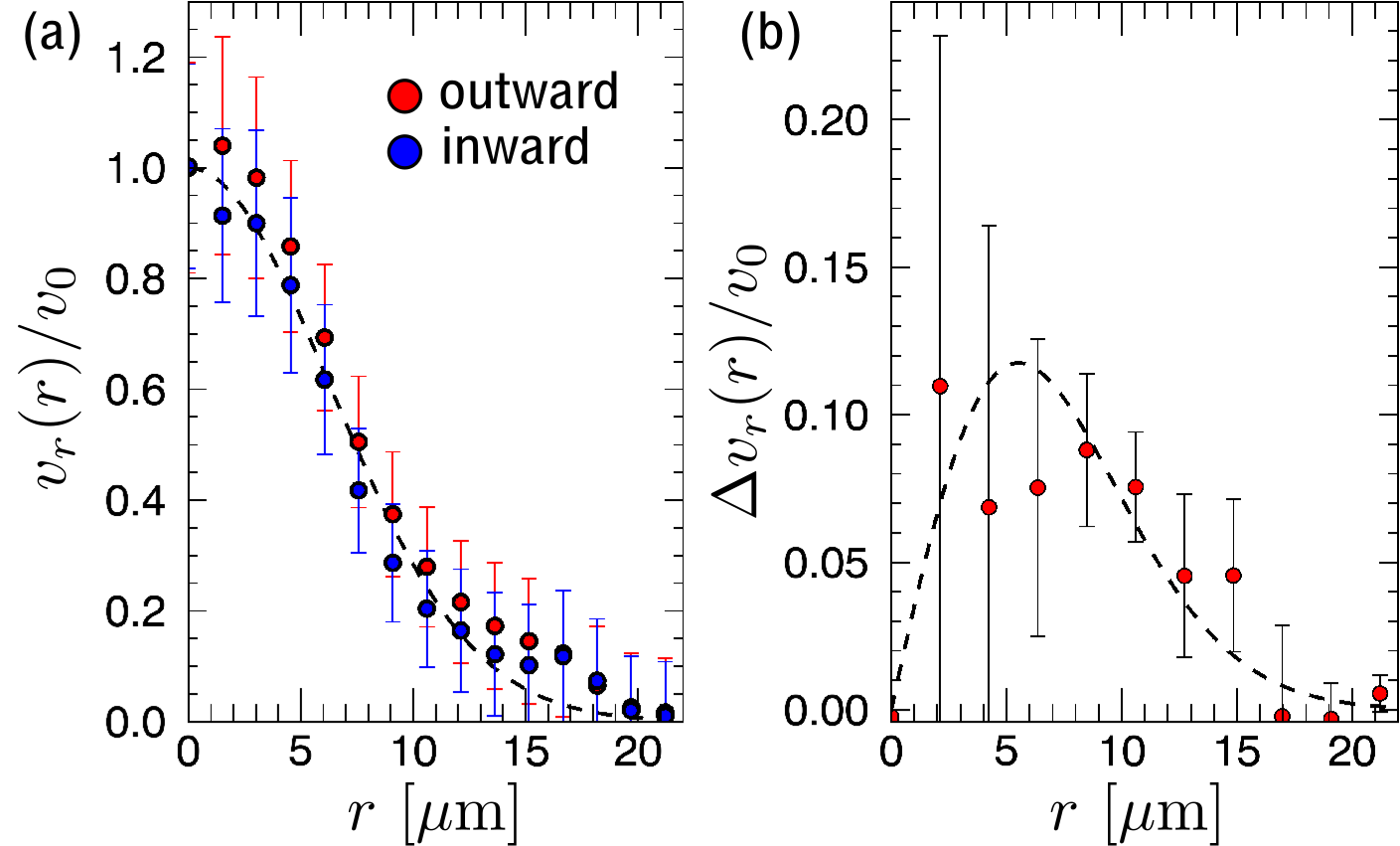}
  \caption{(a) Radial component of the swimmer's velocity, $v_r(r)$, 
    as it moves either inward toward the center of the intensity
    distribution (blue symbols) or outward (red).
    The dashed curve is a fit to a Gaussian
    corresponding to the measured intensity distribution of the
    illumination.
    Speeds are normalized by the peak speed, $v_0$, of this fit.
    (b) Difference, $\Delta v_r(r)$, between outward and inward radial speeds, together
    with a fit to Eq.~\eqref{eq:deltav}.}
  \label{fig:radial}
\end{figure}

Figure~\ref{fig:light_trajectory}(c) establishes, broadly speaking, 
that the swimmer's speed is proportional to the local light intensity.
This intensity distribution, moreover, is reasonably well described
as a Gaussian surface of revolution.
In that case, the swimmer's speed should fall off as a Gaussian
with distance from the optical axis.
This is consistent with the tracking data for the swimmer's
radial speed, plotted in Figure~\ref{fig:radial}, using the
width $\sigma = \SI{7.3}{\um}$ obtained from imaging photometry
of the beam.

These data, however, also show a more subtle feature:
the radial speed is slightly but systematically higher when
the particle is moving radially outward,
$\vec{v}(t) \cdot \hat{r} > 0$, than when it moves
radially inward, $\vec{v}(t) \cdot \hat{r} < 0$.
The radial speeds plotted in Figure~\ref{fig:radial}(a) are computed
separately for the inward- and outward-moving trajectory segments.
For these measurements, the swimmer is defined to be 
located at the position of its cube, $\vec{r}_c$.
This avoids systematic offsets due to variations in
$I(\vec{r})$ across the sphere's diameter.
The difference between inward and outward radial speeds, 
is plotted in Figure~\ref{fig:radial}.

The difference in radial speeds is consistent with the action
of a small outward-directed force.
The dashed curve is a comparison with expectations
for the in-plane component of radiation pressure,
\begin{equation}
  \label{eq:deltav}
  \Delta v_r(r) = \alpha_{r} I(\vec{r}) \frac{r}{\sqrt{r^2 + L^2}},
\end{equation}
where the constant $\alpha_{r}$ sets the scale
for the light-matter coupling.
We already have determined that the influence of
radiation pressure on the TPM sphere is negligibly weak.
Any influence of radiation pressure therefore must
result from absorption of light by the hematite cube,
which also is responsible for the swimmer's self-thermophoresis.
The dashed curve in Figure~\ref{fig:radial}(b) is obtained with
$\alpha_r = \SI{3(2)}{\cubic\um\per\second\per\milli\watt}$
and no other adjustable parameters.
This is consistent with the expected in-plane component of the
radiation pressure that would arise from complete absorption of the light by a cube
\SI{300}{\nm} on a side.

\begin{figure*}
  \centering
  \includegraphics[width=0.65\textwidth]{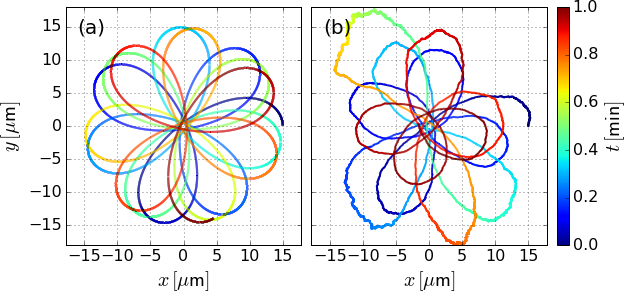}
  \caption{Computed trajectory for a swimmer moving under
    the combined influence of self-thermophoresis and 
    optical torque.  (a) Trajectory traced out in the absence of
    thermal noise.  (b) Equivalent trajectory including both
    translational and rotational diffusion.}
  \label{fig:computedtrajectory}
\end{figure*}

The mean radial velocity of the swimmer thus provides evidence for
a small contribution from radiation pressure acting on the cube
that supplements the hydrodynamic force acting on the sphere.
Although radiation pressure is weak, it strongly influences the
swimmer's motion by exerting a torque on the sphere.
The torque arises because radiation pressure acts on the
cube rather than on the sphere, and acts in the
radial direction, $\hat{r}$, which need not be aligned with the
swimmer's orientation, $\hat{n}$.
Radiation pressure therefore causes rotation about the axis of the sphere
with an angular velocity
\begin{equation}
  \label{eq:angularvelocity}
  \boldsymbol{\omega}(\vec{r})
  =
  \mu_\theta \alpha_r \frac{a}{\sqrt{2}}
  I(\vec{r}) \, \frac{r}{\sqrt{r^2 + L^2}} \,
  \hat{r} \times \hat{n},
\end{equation}
where $\mu_\theta$ is the swimmer's rotational mobility.

\section{Trochoidal motion}
\label{sec:trochoidalmotion}

If, for the sake of simplicity, we consider optical torque but neglect the contribution of
radiation pressure to the swimmer's velocity, and furthermore treat the
sphere as being small ($a \ll \sigma$), we then may model the
velocity as being proportional to the local intensity
\begin{equation}
  \label{eq:simplemodel}
  v(r) \approx \frac{v_0}{I_0} \, I(r)
\end{equation}
with a characteristic speed for the present system
$v_0 = \SI{40(2)}{\um\per\second}$.
The particle also diffuses under the influence
of random thermal forces.
The set of coupled Langevin equations describing 
the particle's in-plane are
\begin{subequations}
\label{eq:eom}
\begin{align}
  \label{eq:xequation}
  \dot{x}(t)
  & =
    v(r(t)) \cos(\phi(t))  + \zeta(t)
    \quad \text{and}\\
  \label{eq:yequation}
  \dot{y}(t)
  & =
    v(r(t)) \sin(\phi(t)) + \zeta(t),
\end{align}
where
$\phi(t) = \tan^{-1}(y(t)/x(t))$
is the swimmer's orientation relative to $\hat{x}$, and
$\zeta(t)$ is a normally distributed random variable
that models thermal noise.
The swimmer's orientation evolves according to
Eq.~\eqref{eq:angularvelocity}, and also is influenced
by thermal fluctuations,
\begin{equation}
  \label{eq:phidot}
  \dot{\phi}(t)
  =
  \frac{x(t) \dot{y}(t) - y(t) \dot{x}(t)}{d \sqrt{r^2(t) + L^2}} +
  \sqrt{\frac{3}{4a^2}} \, \zeta(t),
\end{equation}
\end{subequations}
where $d$ sets the scale of rotations relative to translations.

Figure~\ref{fig:computedtrajectory}(a) shows a typical trajectory
obtained by Euler-Maruyama integration of Eq.~\eqref{eq:eom} in the
absence of noise ($\zeta(t) = 0$) 
for parameters chosen to mimic the
experimental conditions in Figure~\ref{fig:light_trajectory}(b).
These deterministic equations admit periodic solutions,
\begin{subequations}
\begin{align}
  \label{eq:3}
  x(t + T) & = x(t) \\
  y(t + T) & = y(t) \\
  \phi(t + T) & = \phi(t) + 2 m \pi,
\end{align}
\end{subequations}
where $m$ is an integer.
Traces such as the example in Figure~\ref{fig:computedtrajectory}(a)
at least qualitatively resemble rosette curves.

\begin{subequations}
\label{eq:circular}
One special case of a deterministic rosette
trajectory is the circular path,
\begin{align}
  x_0(t)
  & = 
    \frac{v_0}{\omega} \sin(\omega t) \\
  y_0(t)
  & =
    - \frac{v_0}{\omega} \cos(\omega t),
\end{align}
 which is generated by having the swimmer rotate once per orbit:
\begin{equation}
  \phi_0(t) = \omega t.
\end{equation}
This trajectory has a radius $R$ that satisfies
\begin{equation}
  R^2 = \frac{1}{2} \, d^2 \, \left(1 + \sqrt{1 + 4 \frac{L^2}{d^2}}\right)
\end{equation}
and a corresponding angular frequency
\begin{equation}
  \omega = \frac{2 \pi}{T} = \frac{v_0}{R} \, \exp\left(-\frac{R^2}{2\sigma^2}\right).
\end{equation}
\end{subequations}
\begin{subequations}
\label{eq:rosette}
Perturbing this solution,
\begin{equation}
  \phi(t) = \phi_0(t) + \epsilon \sin(b \omega t),
\end{equation}
yields the hypotrochoidal solutions
\begin{align}
  x(t)
  & =
    x_0(t)
    -
    \frac{\epsilon v_0}{2 \omega} \left[
    \frac{\sin\left((1-b) \omega t\right)}{1-b}
    -
    \frac{\sin\left((1+b) \omega t\right)}{1+b}
    \right] \\
  y(t)
  & =
    y_0(t)
    +
    \frac{\epsilon v_0}{2 \omega} \left[
    \frac{\cos\left((1-b) \omega t\right)}{1-b}
    -
    \frac{\cos\left((1+b) \omega t\right)}{1+b}
    \right],
\end{align}
\end{subequations}
up to corrections at $\order{\epsilon^2}$.
These solutions self-consistently satisfy
Eq.~\eqref{eq:eom} up to $\order{\epsilon^2}$
for the particular choice $b = \sqrt{2}$, and
closely resemble the numerical solution
in Figure~\ref{fig:computedtrajectory}(a) for
appropriate choices of the perturbation amplitude, $\epsilon$.

Other solutions include non-repeating trajectories
in which the particle swims radially outward, never to return.
We designate these as open or escape trajectories.

Figure~\ref{fig:computedtrajectory}(b) shows a typical
numerical solution of the trochoidal trajectory 
from Figure~\ref{fig:computedtrajectory}(a) 
including the influence
of random thermal forces and torques at room temperature.
This simulation was performed with $d = \SI{3.57}{\um}$,
which suggests that that the rotation rate due to optical torque
is only a third the rate of rotational diffusion.
Optical torque, however, acts in a sense governed by the swimmer's
orientation, and so imposes a coherent structure on the trajectory.
Small translational displacements roughen the curve but otherwise
have little effect on its extent or orbital frequency.
Random rotations, however, transfer the swimmer from one
trochoidal orbit to another.
Because an orbit's radial extent is strongly influenced by
the swimmer's orientation near the origin, small diffusional
rotations can have a disproportionate influence on the scale of the pattern
that the swimmer traces out.
This process accounts for the ability of the trap to capture
a swimmer, and provides a mechanism for the swimmer
ultimately to escape.

\section{Conclusions}

\begin{figure}[!t]
  \centering
  \includegraphics[width=\columnwidth]{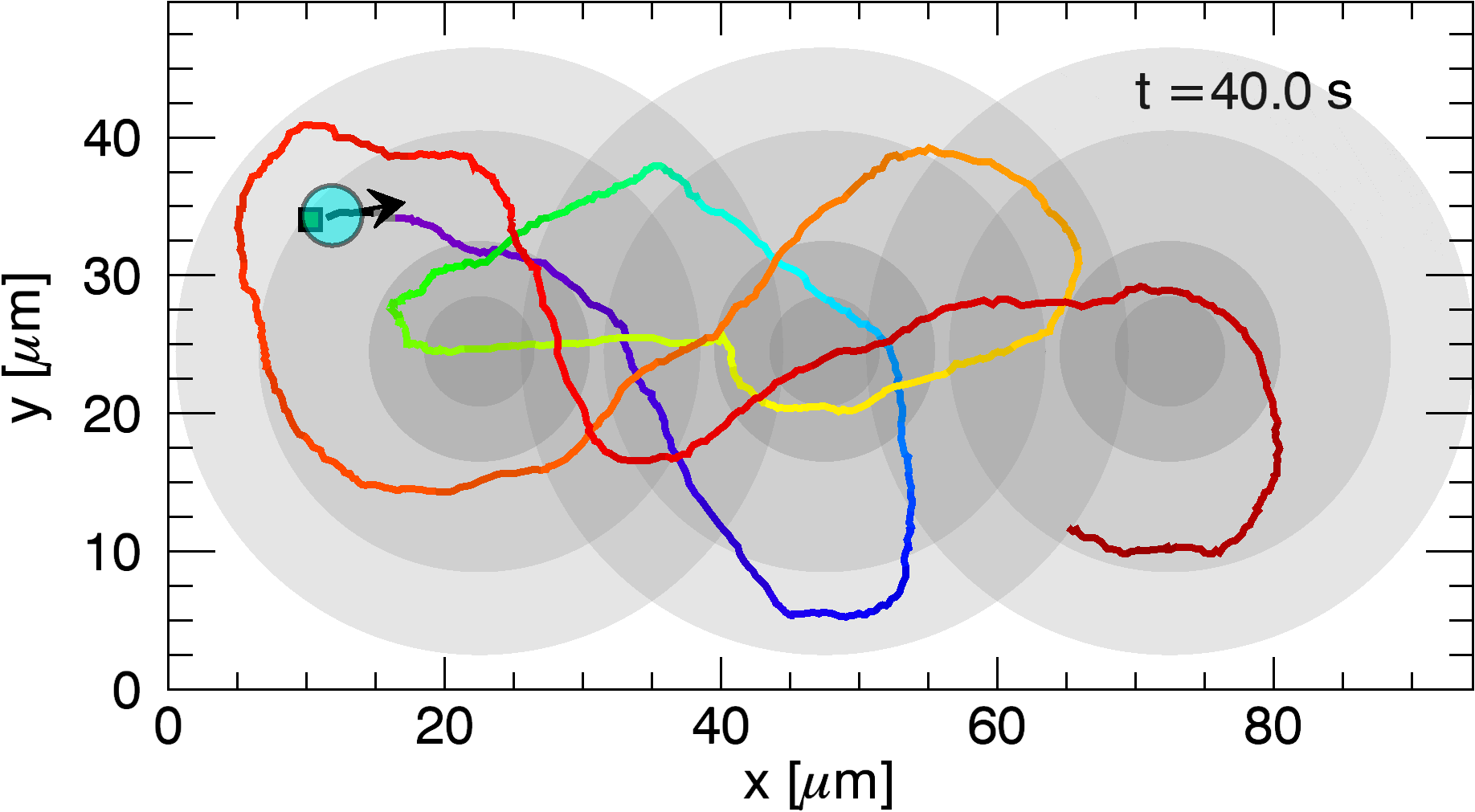}
  \caption{Measured trajectory of a swimmer moving
    under the influence of three
    defocused point traps over the course of \SI{40}{\second}.
    Gray regions are contours of the light
    intensity.  The swimmer is drawn to scale at the outset of its
    trajectory.}
  \label{fig:threetraps}
\end{figure}

Composite colloidal particles consisting of optically absorbing
hematite cubes embedded in transparent colloidal spheres
act as self-thermophoretic swimmers.
These particles move through water when illuminated without requiring
chemical fuel, employing a propulsion mechanism that is captured
by a low-level stokeslet formulation.
When activated by a nonuniform light field, these swimmers also
experience optically-induced torques that tend to orient their
motions.
In the particular case of diverging illumination, this causes the
swimmers to become confined to looping paths that we identify
with rosette curves in general and trochoids in particular.
Random thermal forces and torques perturb these patterns
into paths that we call stochastic trochoidal trajectories.
The observation of such trajectories provides experimental support
for the proposed propulsion mechanism.

Optically-activated self-thermophoretic swimmers constitute a
rich model system for studying active matter.
While the present study focuses on the behavior of a single
particle moving in a comparatively simple light field, prospects
are bright for studying behavior in more sophisticated systems.
Figure~\ref{fig:threetraps}, for example, shows the experimentally
measured trajectory of a single particle moving under the influence
of an array of three uniformly bright defocused optical tweezers, each of which
resembles the single tweezer used for the foregoing studies.
The particle repeatedly loops through the entire pattern.

Multiple swimmers moving through the same light field are
driven both by their interaction with the light and also by their
long-ranged hydrodynamic interactions.  The strength and
direction of these interactions depends on the swimmers'
positions in the light field and on their relative orientations.
Introducing position-dependent coupling in nonuniform light
fields creates an opportunity to study the influence of
dynamic heterogeneity on collective behavior in active matter.

\section*{Acknowledgements}
This work was supported primarily by the National Science Foundation through
Award Number DMR-1305875 and partially by the
MRSEC Program of the National Science Foundation under Award Number DMR-1420073.
Additional support was provided 
by the Gordon and Betty Moore Foundation through Grant GBMF3849.

\footnotesize{

\begin{mcitethebibliography}{26}
\providecommand*{\natexlab}[1]{#1}
\providecommand*{\mciteSetBstSublistMode}[1]{}
\providecommand*{\mciteSetBstMaxWidthForm}[2]{}
\providecommand*{\mciteBstWouldAddEndPuncttrue}
  {\def\EndOfBibitem{\unskip.}}
\providecommand*{\mciteBstWouldAddEndPunctfalse}
  {\let\EndOfBibitem\relax}
\providecommand*{\mciteSetBstMidEndSepPunct}[3]{}
\providecommand*{\mciteSetBstSublistLabelBeginEnd}[3]{}
\providecommand*{\EndOfBibitem}{}
\mciteSetBstSublistMode{f}
\mciteSetBstMaxWidthForm{subitem}
{(\emph{\alph{mcitesubitemcount}})}
\mciteSetBstSublistLabelBeginEnd{\mcitemaxwidthsubitemform\space}
{\relax}{\relax}

\bibitem[Golestanian \emph{et~al.}(2005)Golestanian, Liverpool, and
  Ajdari]{golestanian05}
R.~Golestanian, T.~B. Liverpool and A.~Ajdari, \emph{Phys. Rev. Lett.}, 2005,
  \textbf{94}, 220801\relax
\mciteBstWouldAddEndPuncttrue
\mciteSetBstMidEndSepPunct{\mcitedefaultmidpunct}
{\mcitedefaultendpunct}{\mcitedefaultseppunct}\relax
\EndOfBibitem
\bibitem[Marchetti \emph{et~al.}(2013)Marchetti, Joanny, Ramaswamy, Liverpool,
  Prost, Rao, and Simha]{marchetti13}
M.~C. Marchetti, J.~F. Joanny, S.~Ramaswamy, T.~B. Liverpool, J.~Prost, M.~Rao
  and R.~A. Simha, \emph{Rev. Mod. Phys.}, 2013, \textbf{85}, 1143\relax
\mciteBstWouldAddEndPuncttrue
\mciteSetBstMidEndSepPunct{\mcitedefaultmidpunct}
{\mcitedefaultendpunct}{\mcitedefaultseppunct}\relax
\EndOfBibitem
\bibitem[Walther and M\"uller(2008)]{walther08}
A.~Walther and A.~H.~E. M\"uller, \emph{Soft Matter}, 2008, \textbf{4},
  663--668\relax
\mciteBstWouldAddEndPuncttrue
\mciteSetBstMidEndSepPunct{\mcitedefaultmidpunct}
{\mcitedefaultendpunct}{\mcitedefaultseppunct}\relax
\EndOfBibitem
\bibitem[Paxton \emph{et~al.}(2004)Paxton, Kistler, Olmeda, Sen, St.~Angelo,
  Cao, Mallouk, Lammert, and Crespi]{paxton04}
W.~F. Paxton, K.~C. Kistler, C.~C. Olmeda, A.~Sen, S.~K. St.~Angelo, Y.~Cao,
  T.~E. Mallouk, P.~E. Lammert and V.~H. Crespi, \emph{J. Am. Chem. Soc.},
  2004, \textbf{126}, 13424--13431\relax
\mciteBstWouldAddEndPuncttrue
\mciteSetBstMidEndSepPunct{\mcitedefaultmidpunct}
{\mcitedefaultendpunct}{\mcitedefaultseppunct}\relax
\EndOfBibitem
\bibitem[Jiang \emph{et~al.}(2010)Jiang, Yoshinaga, and Sano]{jiang10}
H.-R. Jiang, N.~Yoshinaga and M.~Sano, \emph{Phys. Rev. Lett.}, 2010,
  \textbf{105}, 268302\relax
\mciteBstWouldAddEndPuncttrue
\mciteSetBstMidEndSepPunct{\mcitedefaultmidpunct}
{\mcitedefaultendpunct}{\mcitedefaultseppunct}\relax
\EndOfBibitem
\bibitem[Volpe \emph{et~al.}(2011)Volpe, Buttinoni, Vogt, K{\"u}mmerer, and
  Bechinger]{volpe11}
G.~Volpe, I.~Buttinoni, D.~Vogt, H.-J. K{\"u}mmerer and C.~Bechinger,
  \emph{Soft Matter}, 2011, \textbf{7}, 8810--8815\relax
\mciteBstWouldAddEndPuncttrue
\mciteSetBstMidEndSepPunct{\mcitedefaultmidpunct}
{\mcitedefaultendpunct}{\mcitedefaultseppunct}\relax
\EndOfBibitem
\bibitem[Buttinoni \emph{et~al.}(2012)Buttinoni, Volpe, K{\"u}mmel, Volpe, and
  Bechinger]{buttinoni12}
I.~Buttinoni, G.~Volpe, F.~K{\"u}mmel, G.~Volpe and C.~Bechinger, \emph{J.
  Phys.: Condens. Matter}, 2012, \textbf{24}, 284129\relax
\mciteBstWouldAddEndPuncttrue
\mciteSetBstMidEndSepPunct{\mcitedefaultmidpunct}
{\mcitedefaultendpunct}{\mcitedefaultseppunct}\relax
\EndOfBibitem
\bibitem[Buttinoni \emph{et~al.}(2013)Buttinoni, Bialk{\'e}, K{\"u}mmel,
  L{\"o}wen, Bechinger, and Speck]{buttinoni13}
I.~Buttinoni, J.~Bialk{\'e}, F.~K{\"u}mmel, H.~L{\"o}wen, C.~Bechinger and
  T.~Speck, \emph{Phys. Rev. Lett.}, 2013, \textbf{110}, 238301\relax
\mciteBstWouldAddEndPuncttrue
\mciteSetBstMidEndSepPunct{\mcitedefaultmidpunct}
{\mcitedefaultendpunct}{\mcitedefaultseppunct}\relax
\EndOfBibitem
\bibitem[Qian \emph{et~al.}(2013)Qian, Montiel, Bregulla, Cichos, and
  Yang]{qian13}
B.~Qian, D.~Montiel, A.~Bregulla, F.~Cichos and H.~Yang, \emph{Chem. Sci.},
  2013, \textbf{4}, 1420--1429\relax
\mciteBstWouldAddEndPuncttrue
\mciteSetBstMidEndSepPunct{\mcitedefaultmidpunct}
{\mcitedefaultendpunct}{\mcitedefaultseppunct}\relax
\EndOfBibitem
\bibitem[Bickel \emph{et~al.}(2014)Bickel, Zecua, and W\"urger]{bickel14}
T.~Bickel, G.~Zecua and A.~W\"urger, \emph{Phys. Rev. E}, 2014, \textbf{89},
  050303(R)\relax
\mciteBstWouldAddEndPuncttrue
\mciteSetBstMidEndSepPunct{\mcitedefaultmidpunct}
{\mcitedefaultendpunct}{\mcitedefaultseppunct}\relax
\EndOfBibitem
\bibitem[Sacanna \emph{et~al.}(2012)Sacanna, Rossi, and Pine]{sacanna12}
S.~Sacanna, L.~Rossi and D.~J. Pine, \emph{J. Am. Chem. Soc.}, 2012,
  \textbf{134}, 6112--6115\relax
\mciteBstWouldAddEndPuncttrue
\mciteSetBstMidEndSepPunct{\mcitedefaultmidpunct}
{\mcitedefaultendpunct}{\mcitedefaultseppunct}\relax
\EndOfBibitem
\bibitem[Abrikosov \emph{et~al.}(2013)Abrikosov, Sacanna, Philipse, and
  Linse]{abrikosov13}
A.~I. Abrikosov, S.~Sacanna, A.~P. Philipse and P.~Linse, \emph{Superlattices
  Microstructures}, 2013, \textbf{9}, 8904--8913\relax
\mciteBstWouldAddEndPuncttrue
\mciteSetBstMidEndSepPunct{\mcitedefaultmidpunct}
{\mcitedefaultendpunct}{\mcitedefaultseppunct}\relax
\EndOfBibitem
\bibitem[Palacci \emph{et~al.}(2013)Palacci, Sacanna, Steinberg, Pine, and
  Chaikin]{palacci13}
J.~Palacci, S.~Sacanna, A.~P. Steinberg, D.~J. Pine and P.~M. Chaikin,
  \emph{Science}, 2013, \textbf{339}, 936--940\relax
\mciteBstWouldAddEndPuncttrue
\mciteSetBstMidEndSepPunct{\mcitedefaultmidpunct}
{\mcitedefaultendpunct}{\mcitedefaultseppunct}\relax
\EndOfBibitem
\bibitem[Palacci \emph{et~al.}(2014)Palacci, Sacanna, Kim, Yi, Pine, and
  Chaikin]{palacci14}
J.~Palacci, S.~Sacanna, S.-H. Kim, G.-R. Yi, D.~J. Pine and P.~M. Chaikin,
  \emph{Phil. Trans. Roy. Soc. A}, 2014, \textbf{372}, 20130372\relax
\mciteBstWouldAddEndPuncttrue
\mciteSetBstMidEndSepPunct{\mcitedefaultmidpunct}
{\mcitedefaultendpunct}{\mcitedefaultseppunct}\relax
\EndOfBibitem
\bibitem[Weinert and Braun(2008)]{weinert08}
F.~W. Weinert and D.~Braun, \emph{Phys. Rev. Lett.}, 2008, \textbf{101},
  168301\relax
\mciteBstWouldAddEndPuncttrue
\mciteSetBstMidEndSepPunct{\mcitedefaultmidpunct}
{\mcitedefaultendpunct}{\mcitedefaultseppunct}\relax
\EndOfBibitem
\bibitem[Yang and Ripoll(2013)]{yang13}
M.~Yang and M.~Ripoll, \emph{Superlattices Microstructures}, 2013, \textbf{9},
  4661--4671\relax
\mciteBstWouldAddEndPuncttrue
\mciteSetBstMidEndSepPunct{\mcitedefaultmidpunct}
{\mcitedefaultendpunct}{\mcitedefaultseppunct}\relax
\EndOfBibitem
\bibitem[Crocker and Grier(1996)]{crocker96}
J.~C. Crocker and D.~G. Grier, \emph{J. Colloid Interface Sci.}, 1996,
  \textbf{179}, 298--310\relax
\mciteBstWouldAddEndPuncttrue
\mciteSetBstMidEndSepPunct{\mcitedefaultmidpunct}
{\mcitedefaultendpunct}{\mcitedefaultseppunct}\relax
\EndOfBibitem
\bibitem[Krishnatreya and Grier(2014)]{krishnatreya14a}
B.~J. Krishnatreya and D.~G. Grier, \emph{Opt. Express}, 2014, \textbf{22},
  12773--12778\relax
\mciteBstWouldAddEndPuncttrue
\mciteSetBstMidEndSepPunct{\mcitedefaultmidpunct}
{\mcitedefaultendpunct}{\mcitedefaultseppunct}\relax
\EndOfBibitem
\bibitem[Silverman(1992)]{silverman92}
B.~W. Silverman, \emph{Density Estimation for Statistics and Data Analysis},
  Chapman \& Hall, New York, 1992\relax
\mciteBstWouldAddEndPuncttrue
\mciteSetBstMidEndSepPunct{\mcitedefaultmidpunct}
{\mcitedefaultendpunct}{\mcitedefaultseppunct}\relax
\EndOfBibitem
\bibitem[Iacopini \emph{et~al.}(2006)Iacopini, Rusconi, and Piazza]{iacopini06}
S.~Iacopini, R.~Rusconi and R.~Piazza, \emph{Euro. Phys. J. E}, 2006,
  \textbf{19}, 59--67\relax
\mciteBstWouldAddEndPuncttrue
\mciteSetBstMidEndSepPunct{\mcitedefaultmidpunct}
{\mcitedefaultendpunct}{\mcitedefaultseppunct}\relax
\EndOfBibitem
\bibitem[Fayolle \emph{et~al.}(2008)Fayolle, Bickel, and W\"urger]{fayolle08}
S.~Fayolle, T.~Bickel and A.~W\"urger, \emph{Phys. Rev. E}, 2008, \textbf{77},
  041404\relax
\mciteBstWouldAddEndPuncttrue
\mciteSetBstMidEndSepPunct{\mcitedefaultmidpunct}
{\mcitedefaultendpunct}{\mcitedefaultseppunct}\relax
\EndOfBibitem
\bibitem[Pozrikidis(1992)]{pozrikidis92}
C.~Pozrikidis, \emph{Boundary Integral and Singularity Methods for Linearized
  Viscous Flow}, Cambridge University Press, New York, 1992\relax
\mciteBstWouldAddEndPuncttrue
\mciteSetBstMidEndSepPunct{\mcitedefaultmidpunct}
{\mcitedefaultendpunct}{\mcitedefaultseppunct}\relax
\EndOfBibitem
\bibitem[Blake(1971)]{blake71}
J.~R. Blake, \emph{Proc. Cambridge Phil. Soc.}, 1971, \textbf{70},
  303--310\relax
\mciteBstWouldAddEndPuncttrue
\mciteSetBstMidEndSepPunct{\mcitedefaultmidpunct}
{\mcitedefaultendpunct}{\mcitedefaultseppunct}\relax
\EndOfBibitem
\bibitem[Liron and Mochon(1976)]{liron76}
N.~Liron and S.~Mochon, \emph{J. Eng. Math.}, 1976, \textbf{10}, 287\relax
\mciteBstWouldAddEndPuncttrue
\mciteSetBstMidEndSepPunct{\mcitedefaultmidpunct}
{\mcitedefaultendpunct}{\mcitedefaultseppunct}\relax
\EndOfBibitem
\bibitem[Dufresne \emph{et~al.}(2001)Dufresne, Altman, and Grier]{dufresne01}
E.~R. Dufresne, D.~Altman and D.~G. Grier, \emph{Europhys. Lett.}, 2001,
  \textbf{53}, 264--270\relax
\mciteBstWouldAddEndPuncttrue
\mciteSetBstMidEndSepPunct{\mcitedefaultmidpunct}
{\mcitedefaultendpunct}{\mcitedefaultseppunct}\relax
\EndOfBibitem
\bibitem[Spagnolie and Lauga(2012)]{spagnolie12}
S.~E. Spagnolie and E.~Lauga, \emph{J. Fluid Mech.}, 2012, \textbf{700},
  105--147\relax
\mciteBstWouldAddEndPuncttrue
\mciteSetBstMidEndSepPunct{\mcitedefaultmidpunct}
{\mcitedefaultendpunct}{\mcitedefaultseppunct}\relax
\EndOfBibitem
\end{mcitethebibliography}
\providecommand*{\mcitethebibliography}{\thebibliography}
\csname @ifundefined\endcsname{endmcitethebibliography}
{\let\endmcitethebibliography\endthebibliography}{}
}

\end{document}